\documentclass{article}

\begin{document}

\title{Large Surveys in Cosmology: The Changing Sociology}

\author{Ofer Lahav, Institute of Astronomy, 
Madingley Road, Cambridge CB3 0HA,
UK\\E-mail: lahav@ast.cam.ac.uk}

\maketitle

{\bf Abstract} 
Galaxy redshift surveys and Cosmic Microwave Background experiments
are undertaken with larger and larger teams, 
in a fashion reminiscent of particle physics experiments 
and the human genome projects.  We discuss
the role of young researchers, the issue of multiple authorship,
and ways to communicate effectively in teams of tens
to hundreds of collaborators.
\footnote{Invited article for `Organizations and Strategies in Astronomy II',
Ed. Andre Heck}

\section{Introduction}

The field of Observational Cosmology is going through an `inflationary
phase'. Galaxy redshift surveys will soon register millions of
galaxies, imaging surveys will soon record Terabytes per observing
night, and the Cosmic Microwave Background experiments will map the
sky at very high resolution and sensitivity.

These technological developments have immediate effects on the
human interaction.  We commonly hear at conferences statements
like ``to make a big impact in Astronomy, a new survey must be 10-1000
times better in sensitivity and/or resolution and/or the number of
objects'', and ``it takes at least 2-3 times the most pessimistic
estimate to begin/complete/analyse a survey''.  This indicates that 
conducting large surveys in this competitive field is becoming more
demanding, and that the new technology heavily relies on human
resources.  While improving the technology of the big surveys, the
human aspects should not be neglected.  This raises some questions about
the changing sociology of doing research in modern Astronomy:

$\bullet$
What is the individual's contribution in a big collaboration 
(cf. the particle physics experiments and the human genome project ) ?

$\bullet$ 
How to communicate effectively (via e-mail, web, meetings)
in a team of 20-200 collaborators ?

$\bullet$
What skills should be acquired by the next generation of
Astronomers ?

$\bullet$
Will the increase in projects and data sets be followed by
more jobs for young astronomers ?   

$\bullet$ 
How should the community deal with public domain data 
(e.g. HDF and the proposed Virtual Observatories) ?  

$\bullet$
How to communicate the
knowledge resulting from the surveys to the tax-payer ?

Here we examine some of these issues.  I happen to be
involved in several large collaborations, and  I have chosen to illustrate
some of the points by using as an example 
 the ongoing 2-degree-Field Galaxy Redshift
Survey (2dFGRS).  Needless to say, the points made below are my own
views, and they do not necessarily represent a `party line'.

\section{Large surveys, large collaborations}

Astronomy has been unique among the sciences
in allowing small teams (of 3-8 people) to compete
for time on world-class telescopes.
Although the telescopes  are built by hundreds of
people, time is allocated (via competition)
among the community at large, and only those involved in
the science appear on the resulting paper (with the facility
acknowledged and the instrument builders referenced). 
This made
Astronomy more attractive to some young researchers than 
say particle physics, where papers include hundreds of authors. 
However, the new big surveys in Astronomy require many more participants.

We shall first discuss the big redshift surveys.  Multi-fibre
technology now allows us to measure redshifts of millions of galaxies.
Two major surveys are underway.  The US Sloan Digital Sky Survey (SDSS
\footnote {http://www.sdss.org/}) will measure redshifts of about 1
million galaxies over a quarter of the sky.  The Anglo-Australian 2
degree Field (2dF) survey
\footnote {http://www.mso.anu.edu.au/2dFGRS/} 
will measure redshifts for 250,000 galaxies
selected from the APM catalogue.  Over 150,000 2dF redshifts have been
measured so far (as of April 2001).  

By the standards of the new millennium the 2dFGRS is a medium-sized
collaboration (20-30 people).  For reference, the QSO 2dF survey
\footnote {http://www.2dfquasar.org/} has only 6 collaborators.  On the
other hand, the ambitious Sloan Digital Sky Survey (SDSS) involves a
formal list of roughly 100 `builders', people who have put at least 2
years effort into the project (York et al. 2000).  The SDSS home page
on collaboration
\footnote
{http://www.sdss.org/collaboration/index.html} mentions 150 scientists
at the 11 participating institutions, but there 
are about 400 people with data rights. 
The SDSS is coordinated by a Collaboration Council
(`CoCo') which helps to manage the affairs of the collaboration.
Examples of other large `ground-based' collaborations are
micro-lensing experiments (e.g. Machos) 
and imaging surveys (e.g. Vista, which involves 18 universities in the UK).

The Cosmic Microwave Background (CMB) experiments also require large
collaborations with new management strategies.  While recent and ongoing 
experiments like Boomerang
\footnote{http://www.physics.ucsb.edu/~boomerang/team.html},
Maxima 
\footnote{http://cfpa.berkeley.edu/group/cmb/maximapeople.html}
and MAP 
\footnote{http://map.gsfc.nasa.gov/html/institutions.html}
have  `only' 20-40  collaborators each, the Planck
project is of a different order of magnitude.

In the Planck project
\footnote{http://astro.estec.esa.nl/Planck/}  (to be launched in 2007) there are several
degrees of involvement (without counting the people involved from 
industry).  There are 50-100 people on the management level, some
100-200 people involved with the instrumentation, and at least 30
people who are involved with data analysis on a day-to-day basis
\footnote{For the big projects
like SDSS and Planck it was actually difficult to find accurate estimates
for the number of people involved. This is by itself
an interesting fact.}.  

When it comes to considering the optimal size of a collaboration, it is
worth recalling some remarks from the autobiography of Fred Hoyle
(1994): ``The essential point - the overriding point- is that the
number of people with whom we need to interact in our daily lives
should not exceed about one hundred, and preferably, on any enterprise
of difficulty, not more than twenty-five.  This is because
twenty-five was the typical size of the hunting parties of
pre-history. It is the scale of the medieval village, the scale of the
modern cabinet in government, ...  More or less everything that lies
within it will be successful, and more or less everything that lies
outside it will not''.

\section{2dFGRS as a test case}

The 2dFGRS (e.g. Folkes et al. 1999; Peacock et al. 2001) includes
about 20 core members  and in addition about 10 students and
post-docs who are heavily involved in the survey.  Most of the
collaborators are in various institutions in the UK and in Australia,
so there are in fact two sub-teams, with principal investigators in
each of the countries.  This geographical distribution has led to
regular `half-team' 1-day meetings (about 2-3 per year) in the two
countries, and to regular Email/WWW exchange.  The 2dFRGS group
web-site has proved useful for exchanging data, results, and minutes of
meetings among the team members.

Since the time the 2dFGRS team formed (around 1995), 
some members have left and others
have joined.  Also, the scientific goals have somewhat changed given
the rapid progress in other areas of Cosmology (e.g. the CMB).  This
requires frequent updates on `who is doing what', with careful
attention to protecting the work of PhD students and post-docs.

The 2dFGRS is not complete yet, so it is too early to assess its
overall performance, but on the whole one can make the following
observations:

* It has taken some time for all involved to get used
to the `loss of individuality' and  
to the structure of the big collaboration, to agree on the division 
of labour, and to develop appropriate communication channels.

* The regular  meetings in the UK and in Australia
are very useful in focusing attention on technical issues
and progress with papers.

* The Email/WWW exchange is quite efficient, even if daunting at times
(see below).

* Decisions on papers and authorship have usually 
reached reasonable agreement after iterations among the relevant people.

* Requests for external collaborators have been  dealt with 
in a democratic way, by consulting the entire team.

\section{The role of PhD students and Post-Docs}

Young researchers may find themselves in big collaborations,
co-authoring papers with tens of authors, and with collaborators
they have never met.
Although there is a danger that individuals (junior as well
as senior) might be `lost in the crowd', there are some benefits 
for a young person to be involved in a collaboration 
at the forefront of research.
However, it is crucial to identify a niche, which is not already taken
by other senior members of the collaborations, or by other students.
When a PhD student or a post-doc has such a territory, his/her work,
if of high quality, is recognized by a large number of people well
ahead of publication.  There is also a constant exchange of ideas and
cross-fertilization with collaborators who are leaders of the field.
This means that the young person can get exposure and can form a
international `network' at an early stage of his/her career.  Being
appreciated by a number of senior people also improves the chances
of getting post-doctoral and faculty positions.

There is however, at least one problematic issue
related to long term projects. 
Some post-docs  are employed for a period of 2-3 years primarily to
develop algorithms and pipe-line software for future experiments.  
This means that when they
next apply for jobs, it would be difficult for them to present `real'
scientific output.  
The `reward system' 
for those who put in several years of hard work
on technical aspects of the surveys
varies from country to country 
and from one institution to another.
Is some places  more can be done
to improve it. 

There are a number of solutions to this problem
(heard occasionally in Cambridge pubs):

* To assign some of the development task to national laboratories, 
where PhDs with permanent positions can develop 
long term projects without the worry of `publish or perish'. 

* To enable post-docs who are working on software development etc.
to spend say 50 \% of their time on science of their choice. 

* To change the rules of assessment for positions from being based
 entirely on listed  journal papers to other products such as
 software packages or management achievements. 
 There should be  career paths
 for such people which are as highly regarded as the standard academic
 university track.
 This assessment would of
 course heavily depend on references from senior members of the
 collaboration.

Another aspect of the large surveys is that the skills required
for some of the tasks are quite different from the PhD qualifications
of the previous generation.
While there is still great need for post-docs and students 
with analytic skills and deep knowledge of the Landau \& Lifshitz 
volumes, we see a new generation of successful post-docs
who have stronger emphasis on numerical and computational work.
The group dynamic of the large collaborations also suggests that 
those with good communication skills have better chances of succeeding. 

It remains to be seen if the growth in projects and `soft-money'
positions will eventually lead to more tenured positions in 
Astronomy. This issue is beyond the scope of this article,
but it is clear that the probability of a young researcher 
eventually getting a permanent position also affects the   
research patterns in the large collaborations.

\section{Email traffic and the WWW}

Joining big teams also means spending a large fraction of the day on
Email.  The Email and the World Wide Web (WWW) 
media make the interaction between people in
different institutions and countries easier, but it consumes time and
energy.  The ethics of using Email have not yet been structured
in the society (see e.g. Wallace 1999),
and one can experience daily different style of Email communications.

In a large collaboration most of the Email messages circulated are
relevant to only a subset of the team.  One may choose to send an
email only to a subset of the team, but then others may get upset
about not being informed !  I found it helpful when a message
circulated to the entire team gives in 1-2 sentences at the top a
summary of the main point, with clear indication of who might be
directly interested in it, and who is expected to act upon it.  It is
also helpful if the sender points to material such as tables and plots
in his/her web home page, instead of sending huge files
to the Email-boxes of numerous collaborators.

In the SDSS collaboration there are different (about 40 in total) 
of Email
exploders for all aspects of SDSS 
(e.g. photometric pipeline, galaxy science, etc.), 
which are archived on the web.
This allows people to choose to pay attention to just those aspects
they find important.

Other problematic issues related to email are well known, e.g.
misunderstanding over language, style and terminology.  
For example, from time to time 
messages with sensitive `political' issues make it (by
chance or by design !) to those who were not supposed to see them.  In
certain circumstances it is worth remembering to use, instead of Email,
the good old telephone !

\section{Multiple Authorship}

We examined the number of authors in volumes of the Astrophysical
Journal (ApJ).  Of the 32 papers published in the first volume of the
ApJ in 1895, 31 (97 \%) were written by a single author.  On the other
hand, in the first volume of ApJ 2000, only 15\% were written by a
single author, 70\% written by 2-5 authors, and 15 \% written by 6 and
more authors.

The recent 2dFGRS papers (e.g. Folkes et al. 1999; Peacock et al.
2001) have about 25-30 authors.  This long list of authors
attempts to reflect the division of labour regarding instrumentation,
observations, data reduction, analysis and theory.  The author-list is
usually led by the 5-6 authors who contributed most to that particular
paper, and the rest are listed by alphabetic order.  The credit for
people is complicated even more by the fact that the big surveys are
stretched over many years, so some participants leave the project (or
quit Astronomy) and others join in.

In the SDSS collaboration there are formal rules about authorship,
e.g. 
a first list of people who did the immediate work on a given paper, and
everybody else alphabetically (similar to 2dFGRS),
but also that no-one is automatically put as a co-author on a paper;
one must explicitly request co-authorship.

Paczynski (2000) raised the question (in the context of monitoring 
the whole sky for variability of objects) ``should the whole effort be combined in a very
large team, with all papers having several dozen authors listed
alphabetically, and no way to find out whom to credit and whom to blame
for different parts of the project ?''.

This problem arises as  the big projects involve individuals who
worked hard on the instrumentation and data reduction, and they
deserve credit for their efforts.  However, one alternative would be to have
technical papers (or web sites) written by those who contributed to
the infra-structure of the project, which will later be quoted by any
paper resulting from the survey.  The same holds for more scientific
aspects of the collaboration, i.e. apart from core papers,
to break the publications down into specific
studies with the authorship of those who directly contributed.
In the case that two (or more) groups within the team
attempt to address the same question by analysing the data differently,
it seems most logical to simply publish two separate papers. 
Another possible  solution for large
collaborative papers is to have authorship by section as well as a global
author list of the paper. This would allow a more precise assignment of
credit/blame to be apportioned.
It will no doubt take some time for the Astronomical community to
develop `rules' regarding authorship and publications.

\section{Public Release, Virtual Observatories and Data Mining}

Time allocation 
committees oblige survey teams 
to release the data 
within a given period.
Perhaps the most successful example is that of the Hubble Deep Field
(Williams et al. 1996),
 where the data were made available to all over the WWW, and
resulted in a remarkable scientific output by groups not necessarily
involved in conducting the survey.  We shall no doubt see similar
trends with future publically available data sets such as 
2dF, SDSS, 2MASS
\footnote{http://www.ipac.caltech.edu/2mass/}
and 6dF 
\footnote{http://www.mso.anu.edu.au/~colless/6dF/}.
There are in fact plans 
to establish Virtual Observatories 
and Astro-grid (distributed CPU) facilities. 
This is an interesting  concept   
where the data produced 
by large teams go eventually to the 
individual, and allow  small groups
to do their own data-mining and analysis.
The exact nature of these new digital research facilities still  
needs to be defined (e.g. Heck 2001).
Another aspect of these public domain data 
is of course that anyone else in world, not only professional 
Astronomers, can access the data, or at least enjoy some pretty pictures.

\section{Discussion}

In recent years the astronomical community has
experienced an enormous growth 
in the number of projects and photons collected by ground-based
and space observatories. 
This has led to a new style of work in large teams, 
and competition between big projects.
As discussed above, these trends could be positive if 
the large projects are divided into smaller tasks that allow
the individuals (in particular young researchers) to identify their niche.
The Astronomical community will have to define  
rules and ethics related to authorship of papers, and Email    
and other communication channels in collaborations of tens to hundreds of
people. While these issues are occasionally discussed informally,
a more open and frank discussion could help to shape the sociology of the 
new Astronomy.

\section*{Acknowledgments}

I thank A. Heck
for suggesting to me to write this article, 
to S. Bridle, A. Bunker, M. Colless, M. Diemling, 
F. van Leeuwen, J. Nicholas and M. Strauss for providing information
and suggestions,
and to the participants of the round table tea at the IoA
for helpful discussions.


\end{document}